\documentclass[journal,12pt,onecolumn,draftclsnofoot,]{IEEEtran}

\usepackage{acronym}
\usepackage{amsmath}
\usepackage{amssymb}
\usepackage{authblk}

\usepackage{bbm}

\usepackage{cite}

\usepackage{dsfont}

\usepackage{graphicx}

\usepackage{mathtools}

\usepackage{physics}

\usepackage{subcaption}

\usepackage{tikz}
\usetikzlibrary{automata, arrows, positioning}
\usepackage{tikz-network}

\newacro{BSM}{Bell State Measurements}
\newacro{CNOT}{Controlled-$X$}
\newacro{GHZ}{Greenberger-Horne-Zeillinger}

\newcommand{\hilb}[0]{%
  \mathcal{H}%
}

\title{Universal Quantum Walk Control Plane for Quantum Networks}

\author[1]{Matheus Guedes de Andrade}
\author[1, 2]{Nitish K. Panigrahy}
\author[1]{Wenhan Dai}
\author[3]{Saikat Guha}
\author[1]{Don Towsley}
\affil[1]{College of Information and Computer Science, University of Massachusetts}
\affil[2]{School of Engineering and Applied Science, Yale University}
\affil[3]{College of Optical Sciences, University of Arizona}
\date{May 2023}

\begin{document}

\maketitle

\begin{abstract}
    \footnote{A preliminary version of this work was presented at IEEE International Conference on Quantum Computing and Engineering 2021 (QCE21)~\cite{de2021quantum}.
    }
    Quantum networks are complex systems formed by the interaction among quantum processors through quantum channels.
    Analogous to classical computer networks, quantum networks allow for the distribution of quantum operations among quantum processors.
    In this work, we describe a Quantum Walk Control Protocol (QWCP) to perform distributed quantum operations in a quantum network.
    We consider a generalization of the discrete-time coined quantum walk model that accounts for the interaction between quantum walks in the network graph with quantum registers inside the network nodes.
    QWCP allows for the implementation of networked quantum services, such as distributed quantum computing and entanglement distribution, abstracting hardware implementation and the transmission of quantum information through channels.
    Multiple interacting quantum walks can be used to propagate entangled control signals across the network in parallel.
    We demonstrate how to use QWCP to perform distributed multi-qubit controlled gates, which shows the universality of the protocol for distributed quantum computing.
    Furthermore, we apply the QWCP to the task of entanglement distribution in a quantum network.
\end{abstract}
\section{Introduction}\label{sec:introduction}

Quantum networking is an innovative, multidisciplinary field of research that promises revolutionary improvements in communications, enabling tasks and applications that are impossible to achieve with the exclusive exchange of classical information \cite{kimble2008quantum, wehner2018quantum}. Similar to a classical computer network, a quantum network is a distributed system composed of quantum computers and quantum repeaters that exchange quantum information across physical channels. Among applications supported by quantum networks, \textit{Distributed Quantum Computing} (DQC) is of particular interest as it leverages the power of interconnected quantum computers to create a virtual quantum machine with processing capabilities that surpass its physical constituents alone \cite{beals2013efficient, cacciapuoti2019quantum, gyongyosi2021scalable}.
DQC is general as it encompasses any distributed architecture for quantum computing, from the short-range interconnection of multiple \textit{Quantum Processing Units} (QPUs) inside the same cooling device, to arbitrary-scale quantum data centers~\cite{bravyi2022future}.
Furthermore, DQC offers a practical way to scale quantum computers in the \textit{Noise Intermediate Scale Quantum machines} (NISQ) era~\cite{gyongyosi2021scalable}. Modular, distributed architectures facilitate the construction of large quantum machines by replacing the complexity of building a monolithic QPU with a large number of qubits with that of a machine consisting of an interconnection of simpler QPUs with fewer qubits~\cite{van2016path, ang2022architectures, bravyi2022future}.

When the quantum network scenario is considered, the complexity of distributed quantum computing extends in at least two dimensions. First, physical quantum channels have a well known depletion effect in the exchange of quantum data, e.g., an exponential decrease in channel entanglement rate with distance~\cite{pirandola2017fundamental}. Second, there is a demand for a quantum network protocol capable of performing desired distributed quantum operations while accounting for network connectivity. Generic quantum computation with qubits in distinct quantum processors demands either the application of remote controlled gates~\cite{chou2018deterministic} or the continuous exchange of quantum information.
For both cases, a network protocol is necessary to orchestrate the communication between nodes that are not directly connected with one another.


One challenge in the design of a control protocol is the need for it to be agnostic to hardware implementations. There is a plethora of physical systems suited for quantum computation under investigation, superconducting qubits~\cite{arute2019quantum}, trapped ions~\cite{kielpinski2002architecture, pino2021demonstration} and Silicon-vacancy color centers in diamond~\cite{stas2021high, wang2021robust} to name a few. In addition, there is a diverse investigation in the architectural description of quantum interconnecting devices capable of exchanging quantum information encoded in distinct quantum physical quantities~\cite{awschalom2021development}. This diverse ecosystem of quantum network technologies indicates that distributed quantum computing network protocols need to abstract physical implementations of quantum switches and network connectivity while maintaining universality requirements.

The goal of this article is to propose a control protocol for distributed quantum computing based on discrete-time coined quantum walks \cite{aharonov2001quantum}. Quantum walks are universal for quantum computing \cite{childs2009universal, childs2013universal, lovett2010universal} and have been successfully employed in the quantum network scenario to perform perfect state transfer (PST) between network nodes \cite{nitsche2016walk, zhan2014perfect}, teleportation \cite{li2019new, wang2017generalized} and quantum key distribution (QKD)~\cite{vlachou2018quantum}. Previous works describe ways of distributing entanglement between nodes on a quantum network using the coin space of the walker to propel entanglement generation between qubits \cite{li2019new, zhan2014perfect}. In addition, the formalism of quantum walks with multiple coins enabled the description of an entanglement routing protocol, which interprets qubits within a network node as vertices of an abstract graph used by a quantum walker to generate entanglement~\cite{li2021entangled}. In spite of their relevance, the quantum walk approaches present in the literature are suited to particular network structures and quantum operations. In particular, they consider the case of regular lattices, describing walker dynamics on regular structures and do not address how the quantum walk can be used to perform generic distributed quantum operations. In this context, our work adds to the literature of both quantum walks and quantum networks with a description of a quantum network control protocol that can be applied to arbitrary graphs and perform universal quantum computing in a quantum network.


\subsection{Contributions}

The contributions of this article are three-fold.
\begin{itemize}
    \item We propose a Quantum Walk Control Protocol (QWCP) for distributed quantum computing in a quantum network. The protocol uses quantum walks to propagate control signals to perform quantum operations among physically separated quantum processors. We assume that each processor dedicates part of its internal quantum register to represent walker control signals and describe how the control subsystem interacts with the data subsystem. Data-control interaction is specified by unitary operations that nodes need to implement in order to realize the quantum walk control plane. We specify operations for the interaction of multiple quantum walks that generates entanglement among quantum control signals and allows for distributed execution of complex multi-qubit gates.
    \item We apply the QWCP to distribute controlled quantum gates. We show the universality of the protocol by describing how an arbitrary 2-qubit operation between qubits in distinct nodes of the network is performed with the quantum walk control plane. The protocol is universal in the sense that it implements a universal gate set on the Hilbert space formed by all qubits in the data subsystem of the nodes. Furthermore, it is generic in the sense that it abstracts hardware implementation and channel transmissions, while being well-defined for any network topology. Moreover, we apply the protocol to parallel propagation of quantum control signals through multiple network paths and trees.
    \item We apply the QWCP to perform entanglement distribution in the data subsystem of the nodes. 
\end{itemize}
We assume that quantum error correction is provided by the network, such that nodes contain sufficient number of qubits to implement the required operations fault tolerantly, and describe the control plane in the setting of noiseless, logical qubits. Throughout the remainder of this work, we refer to noiseless qubits and logical qubits interchangeably.

The remainder of this article is structured as follows. In Section~\ref{sec:background}, we present the system model considered in this work, together with the mathematical background needed for the description of the QWCP. We describe the operations in which the QWCP is built upon in Section~\ref{sec:walk}. The quantum operators for the implementation of distributed gates with a quantum walk in the network are described in Section~\ref{sec:propagation}, where the universality of the protocol is demonstrated. The protocol is applied to perform parallel control propagation with multiple walkers in Section~\ref{sec:parallel}. In Section~\ref{sec:distribution}, we apply the QWCP to realize entanglement distribution. Finally, the manuscript is concluded in Section~\ref{sec:conclusion}.

\section{System Model}\label{sec:background}


Consider a symmetric directed graph $G = (V, E)$, with $V$ and $E$ representing the nodes and the edges of the graph, respectively.
$G$ being symmetric implies that $(v, u) \in E$ if, and only if, $(u, v) \in E$.
Let $\delta(v)$ denote the set of neighbors of vertex $v \in V$ and $d(v) = |\delta(v)|$ denote the degree of $v$.
Let $\Delta_p(u, v)$ denote the hop-distance between $u$ and $v$ through a path $p \in G$, and $\Delta(u, v) = \min_p \Delta_p(u, v)$ denote the minimum hop-distance between the nodes.
Throughout this work we refer to the inverse of a binary string $x \in \{0, 1\}^{*}$ as $\overline{x}$, e.g., if $x = 101$, then $\overline{x} = 010$.
A quantum network is a set of quantum hosts (quantum processors) interconnected by a set of quantum channels that allow the exchange of quantum information \cite{wehner2018quantum}.
A host is either a quantum repeater, a quantum router, or a quantum computer with a fixed number of qubits, which performs generic quantum operations.
We represent a quantum network as a symmetric directed graph $N = (V, E)$.
Each node $v \in V$ represents a quantum host that has a set $\mathcal{N}_v$ of qubits that are used to exchange quantum information with nodes in its neighborhood $\delta(v)$ through a set of quantum channels and a set $\mathcal{M}_v$ of qubits that can be processed together at any time.
More precisely, each edge $(u, v) \in E$ represents a quantum channel connecting the qubits in $\mathcal{N}_u$ and $\mathcal{N}_v$ which can interact through operations mediated via the channel.
We will refer to $\mathcal{N}_v$ and $\mathcal{M}_v$ as the networking and data registers of node $v$, respectively. The sets $\mathcal{N} = \bigcup_v \mathcal{N}_v$ and $\mathcal{M} = \bigcup_v \mathcal{M}_v$ are respectively referred to as the \textit{network control plane} and the \textit{network data plane}.
The Hilbert spaces formed by $\mathcal{N}$ and $\mathcal{M}$ are $\mathcal{H}_\mathcal{N}$ and $\mathcal{H}_\mathcal{M}$, respectively.
Let $\hilb_\mathcal{G} = \hilb_\mathcal{N} \otimes \hilb_\mathcal{M}$ denote the joint Hilbert space spanned by the control and data planes.

This network model divides the qubit registers in the nodes into control and data registers.
Such separation is not required for distributed quantum computing, although modeling network control and data planes separately is useful for the design of a quantum network control plane protocol, as has been shown for the classical case with software defined networks (SDNs)~\cite{kreutz2014software}.
This modular architecture decouples network control and computing operations, and enables the protocol to operate in hardware heterogeneous networks where nodes may use distinct physical platforms for qubits, such as trapped ions or color centers, as long as the control operations are implemented.

\subsection{Quantum network protocols}

Consider the system formed by two quantum processors $u$ and $v$ connected by a channel and their respective qubits. A \textit{local operation} is a quantum transformation represented as a separable operator of the form
$O_u \otimes O_v$,
where $O_u$ and $O_v$ acts on the state space of the qubits at processors $u$ and $v$, respectively. A \textit{Local Operation assisted by two-way Classical Communication} (LOCC) is a local operation that depends on classical information exchanged between nodes \cite{pirandola2017fundamental}, \textit{e.g} the unitaries of quantum teleportation. The exchange of classical information is used to apply adaptive operations to qubits in different nodes.
For instance, in teleportation, the classical output of the \textit{Bell State Measurement} (BSM) performed by the sender is transmitted to the receiver, where Pauli operators that depend on the measurements results are applied.

A quantum network protocol for $N$ is an algorithm that operates on the qubits at the nodes, transforming their joint state.
Network protocols can be modelled by LOCCs under the assumption that a sufficient number of maximally entangled states are shared among nodes before hand and LOCCs are used to transform the state of qubits in both control and data planes.
Let $\mathcal{D}_\mathcal{G} = \hilb_{\mathcal{G}} \times \hilb_{\mathcal{G}}$ denote the space of density matrices spanned by all of the qubits in the network.
Superoperators are linear operators that transform density matrices.
A protocol that requires $t$ time steps to complete can be modeled through a superoperator $\Lambda_t: \mathcal{D}_{\mathcal{G}} \to \mathcal{D}_{\mathcal{G}}$ that represents an LOCC as
\begin{align}
    \rho(t) = \Lambda_t[\rho(0)],\label{eq:locc_preshared}
\end{align}
where $\rho: \mathbb{Z} \to \mathcal{D}_\mathcal{G}$ is a time-dependent density operator determining the state of all qubits in the network, and $\rho(0)$ is a density matrix that contains all the pre-shared entanglement. Each step of the protocol is itself an LOCC represented by a superoperator $\Xi_j: \mathcal{D}_\mathcal{G} \to \mathcal{D}_\mathcal{G}$ such that $\Lambda_t = \Xi_{t - 1} \circ \ldots \circ \Xi_0 [\rho(0)]$~\cite{pirandola2019end}, where $\circ$ denotes the composition operation for superoperators. In fact, it is also possible to model a protocol as a sequence of external (mediated by channels) and internal (local to vertices) time-dependent superoperators $\Gamma_t: \mathcal{D}_\mathcal{G} \to \mathcal{D}_\mathcal{G}$ and $\Phi_t: \mathcal{D}_\mathcal{G} \to \mathcal{D}_\mathcal{G}$, respectively, such that the state of the network is described as
\begin{align}
    & \rho(t + 1) = \Phi_{t}[\Gamma_{t}[\rho(t)]],\label{eq:protocol_model}
\end{align}
where $\Gamma_{t}$ does not represent an LOCC, although $\Phi_{t}$ does, and pre-shared entanglement is not assumed.
As an example, it is usually the case for entanglement distribution protocols that $\Gamma_t$ represents link-level entanglement generation protocols performed at all channels of the network and $\Phi_t$ represents either entanglement swap operations or GHZ projections performed at multiple nodes. The pre-shared entanglement model summarized by~\eqref{eq:locc_preshared} has proven useful since it allows for the derivation of fundamental bounds for entanglement distribution rate \cite{pirandola2019end}. In this article, however, our focus is on the design of network protocols for the transmission of quantum information under the assumption that quantum error correction is provided by the network, i.e., the network is noiseless. In particular, we exploit the representation in \eqref{eq:protocol_model} considering $\Gamma_t$ and $\Phi_t$ as fault tolerant operators provided by the network instead of generic superoperators and perform the analysis in the state vector formalism.

\subsection{Quantum walks on graphs}

There are many ways to define a quantum walk on a graph and this article focuses on the discrete-time coined quantum walk model. Given a symmetric directed graph $G = (V, E)$, a coined quantum walk on $G$ is a process of unitary evolution on the Hilbert space $\mathcal{H}_G = \mathcal{H}_{V} \otimes \mathcal{H}_\mathcal{C}$ formed by the edges of the graph, where $\mathcal{H}_{V}$ codifies vertices and $\mathcal{H}_\mathcal{C}$ is the \textit{coin space} of the walker codifying the degrees of freedom the walker can move on.
In particular, each vertex $v$ defines a set of degrees of freedom, or coin values, $\mathcal{C}_v = \{0, 1, \ldots, d(v) - 1\}$ such that, every $(v, u) \in E$ is uniquely assigned to a coin value $c_{vu} \in \mathcal{C}_v$.
In turn, every $(v, u) \in E$ defines a basis vector $\ket{v, c_{vu}}$ for $\mathcal{H}_G$.
The tensor product structure of $\mathcal{H}_G$ imposes that the coin space has $\max_v \delta(v)$ dimensions and the number of basis vectors in $\mathcal{H}_{G}$ exceeds the number of edges in $G$ for non-regular graphs, i.e., graphs with nodes of different degrees.
Later, $c_{v}$ will be used to represent the self-loop $(v, v)$.
The generic state of the walker $\ket{\Psi(t)} = \sum \psi(v, c, t) \ket{v, c}$ is a superposition of the edges of $G$ and the walker evolution is defined as
\begin{align}
    & \ket{\Psi(t + 1)} = S(t)C(t) \ket{\Psi(t)} \label{eq:qwalk},
\end{align}
where $C$ and $S$ are respectively referred to as the coin and shift operators. The coin is a unitary operator of the form
\begin{align}
    & C(t) = \sum_{v} \dyad{v} \otimes C_v(t), \label{eq:walk_coin}
\end{align}
where $C_v: \mathcal{H}_{\mathcal{C}_v} \to \mathcal{H}_{\mathcal{C}_v}$.
The shift can be defined as any permutation operator on the edges of the graph that maps an edge $(v_1, u)$ to an edge $(u, v_2)$. This mapping of edges represents a permutation between states $\ket{v_1, c_{v_1u}}$ and $\ket{u, c_{uv_2}}$. Two shift operators are used throughout this work: the identity operator, which is a trivial permutation of the edges, and the flip-flop shift operator given by
\begin{align}
    & S_f = \sum_{v \in V}\sum_{u \in \delta(v)} \dyad*{v, c_{vu}}{u, c_{uv}} \label{eq:flipflop}, 
\end{align}
which applies, for every $(v, u) \in E$, the permutation $(v, u) \to (u, v)$. $S_f$ reverses edges in the walker wavefunction and is well defined for every symmetric directed graph. In the general case of non-regular graphs, we define coin and shift operators without specifying the transformation of basis vectors that do not correspond to proper edges of the graph and define such transformation as the identity mapping. The labels of nodes and degrees of freedom are expressed as binary strings, and we denote the bit-wise negation of label $c$ as $\overline{c}$. Note that $c_{vu}$ is not necessarily $\overline{c_{uv}}$.

It is of interest to consider systems formed by multiple quantum walks that are allowed to interact with each other.
For simplicity, we consider systems where the coin and shift operators do not introduce entanglement among the states of the different walks, and define a walk-interaction operator to account for such entanglement.
In this case, the coin and shift operators for a single walker system extend to multiple quantum walks by considering direct tensor product extensions.
In particular, consider a system formed by $k$ quantum walks.
Let $C^{j}(t)$ and $S^{j}(t)$ denote the coin and shift operators for the $j$-th walker at instant $t$, for $j \in \{0, 1,\ldots, k - 1\}$, respectively.
The coin operator assumes the form 
\begin{align}
    & C(t) = \bigotimes_{j = 0}^{k - 1}C^{j}(t) \label{eq:sep_coin},
\end{align}
where $C^{j}(t)$ follows \eqref{eq:walk_coin}, and the shift operator is given by
\begin{align}
    & S(t) = \bigotimes_{j = 0}^{k - 1}S^{j}(t) \label{eq:sep_shift},
\end{align}
where $S^{j}$ follows \eqref{eq:flipflop}.
Moreover, let $I(t): \mathcal{H}_G^{k} \to \mathcal{H}_G^{k}$ denote the operator defining the interaction for a system of $k$ quantum walks at instant $t$. We consider operators of the form
\begin{align}
    I(t) = (\sum_{v} \dyad*{v}^{\otimes k} \otimes U_v(t)) + (\mathds{1}_{\mathcal{H}_v^{k}} -  \sum_{v} \dyad*{v}^{\otimes k}) \otimes \mathds{1}_{\mathcal{H}_C^{k}} \label{eq:walk_interaction}
\end{align}
where $U_v(t): \mathcal{H}_C^{k} \to \mathcal{H}_C^{k}$ is an arbitrary unitary operator acting on the coin space of all $k$ walks.
Operators following \eqref{eq:walk_interaction} allow for the interaction of quantum walker systems based on vertex position.
More precisely, two quantum walks on $G$ are allowed to interact non-trivially at time $t$ if, and only if, edges of the form $(v, u) \in E$  for at least one vertex $v \in V$ have non-zero components in the state of both walks at $t$.
Once the interaction operator is considered, the evolution of the system of multiple quantum walks assumes the form
\begin{equation}
    \ket{\Psi(t + 1)} = S(t) C(t) I(t) \ket{\Psi(t)}.\label{eq:multi_qwalk}
\end{equation}
The interaction operator in \eqref{eq:walk_interaction} does not allow quantum walks to propagate among neighboring nodes of the graph, a behavior that is captured exclusively by the shift operator in the evolution of both single and multiple quantum walks. Later on, we will describe walk evolution processes where multiple interaction and coin operators are applied to the system in any order, at the same time step $t$, before the shift operator is applied to the system.

\section{Quantum Walk Control Plane Protocol} \label{sec:walk}

A direct way of controlling operations in a quantum network with a quantum walk is to couple the walker and the qubits in the nodes, using supersposition in the walker system to implement distributed quantum operations. We describe this joint system in the noiseless setting, under the assumption that quantum error correction is provided by the network.
Let $N' = (V, E  \cup \{(v, v), \forall v \in V\})$ be the graph obtained by adding self-loops to a network $N$.
From now on, $E$ denotes the set of network edges that include self-loops.
$\mathcal{H}_W = \mathcal{H}_V \otimes \mathcal{H}_{\mathcal{C}}$ denotes the space of a walker system on $N'$.
$\mathcal{H}_g = \mathcal{H}_W \otimes \mathcal{H}_\mathcal{M} \subseteq \mathcal{H}_{\mathcal{G}}$ denotes the Hilbert space spanned by the walker and the data plane.
The walker system is assumed to be implemented in the network control plane in a distributed way, such that $\mathcal{H}_W \subseteq \mathcal{H}_{N}$.
In essence, each node $v$ contributes some or all of the qubits in $\mathcal{N}_v$ to describe the space $\mathcal{H}_W$.
Since the Hilbert space of the walker system represents an edge of $N'$ as a basis vector, the dimension of the Hilbert space $\mathcal{H}_\mathcal{N}$ must be at least $|E|$ for $\mathcal{H}_W$ to be a subspace.
When $k$ quantum walks are considered, $\mathcal{H}_W^{k}$ is the space formed by all walkers, the dimension of $\mathcal{H}_{N}$ must be at least $|E|^{k}$, and $\mathcal{H}_g = \mathcal{H}_{W}^{k} \otimes \hilb_\mathcal{M} \subseteq \hilb_\mathcal{G}$.
In this context, the size of the logical space required for the network control plane to implement $k$ quantum walkers is on the order of $\mathcal{O}(k \log(|E|))$ qubits.
We specify that each network node contains $\mathcal{O}(k \log|E|)$ control qubits in order to support $k$ concurrent walks, leading to a total of $\mathcal{O}(k |E| \log|E|)$ logical qubits in the control plane.
This conservative approach ensures that the number of control plane qubits is vastly greater than the minimum number required, while defining resources that are polynomial with respect to the network size.
Note that implementing a logical space of this dimension with quantum error correction requires a larger number of physical qubits~\cite{laflamme1997error}, although addressing this requirements is out of the scope of this work.
Throughout the analysis presented, let $\ket{\Psi(t)} \in \mathcal{H}_g$ denote the joint state between the quantum walkers and the data plane.

Quantum walks are useful for implementing distributed quantum control since they naturally capture the notion of locality through connectivity.
The actions of coin and shift operators impose \textit{neighbor locality} on the state of a quantum walk: an edge $(u_1, v)$ can have a non-zero component in the walk state at instant $t$ if, and only if, an edge $(v, u_2)$ has a non-zero component in the state at time $t - 1$.
This property ensures that control signals, i.e. quantum walks, must propagate among neighboring nodes in the network and rules out non-local operations.

In the remainder of this section, we define the primitive operations for the Quantum Walk Control Protocol (QWCP). In practice, implementing QWCP means implementing the primitive operations described therein.

\subsection{Quantum walk initialization}

The ability to initialize quantum walks in the control plane while respecting locality is a fundamental operation needed by the QWCP.
We require that $v$ initialize a quantum walk in states of the form $\ket{v, c}$ for arbitrary $c \in \mathcal{C}_v$ with LOCCs alone.
Since the quantum walk is assumed to have a distributed implementation in the network, the qubits spanning the quantum walk space are spread across network nodes.
LOCCs ensure that any state of such form does not require entanglement across the nodes in order to initialize the walk.
Moreover, it allows for the exchange of classical information among nodes to coordinate the necessary local quantum operations to be performed.

\subsection{Quantum walk operators}

A quantum walk relies on two fundamental operators: the coin and shift operators. We extend the definition of both operators to the joint Hilbert space $\mathcal{H}_g$ of the network control and data planes. The coin in \eqref{eq:walk_coin} assumes the form
\begin{align}
    & C(t) = \sum_{v} \dyad*{v} \otimes C_v(t) \bigotimes_{u \neq v} \otimes \mathds{1}_{\mathcal{M}} \label{eq:extended_coin},
\end{align}
where $\mathds{1}_{\mathcal{M}}$ denotes the identity operator on the data plane of the network. Similarly, the extended shift operator has the form
\begin{align}
    & S = S_f \otimes \mathds{1}_{\mathcal{M}} \label{eq:extended_shift},
\end{align}
where $S_f$ is given by \eqref{eq:flipflop}.
The time-dependent shift operator $S(t)$ either follows \eqref{eq:extended_shift}, or is the identity operator on the space of all qubits in the network.  
Equipping the QWCP with the trivial identity shift is useful as it allows for control signals to be kept within a node as time progresses, i.e., a wait instruction for the control signal.

Enabling QWCP to allow for parallel propagation of multiple control signals is key to efficiently distribute network services.
To this purpose, we allow multiple concurrent quantum walks in the control space $\mathcal{H}_W$ when required.
The coin and shift operators for multiple walkers follow \eqref{eq:sep_coin} and \eqref{eq:sep_shift}, respectively, and do not introduce entanglement among the states of the different walks.
Similar to the single-walker case, both operators are defined for $\mathcal{H}_g$ through an extension with $\mathds{1}_{\mathcal{M}}$. 

\subsection{Control-data interactions}

Interactions between control and data qubits must respect neighbor locality in order for QWCP to be compatible with the quantum walk formalism. Thus, we define such interactions based on the following principle.
The quantum walk control plane can interact with data qubits in a node $v$ at instant $t$ if and only if there exists at least one edge of form $(v, u) \in E$ with non-zero component in $\ket{\Psi(t)}$.
Following this principle, the interaction between a walk and the data qubits in the network has the form
\begin{align}
    & I_{W\mathcal{M}}(t) = \sum_{v \in V} \dyad*{v} \otimes U_v(t) \bigotimes_{u \neq v} \mathds{1}_{\mathcal{M}_u},\label{eq:local_op}
\end{align}
where $U_v(t): (\mathcal{H}_C \otimes \mathcal{H}_{\mathcal{M}_v}) \to (\mathcal{H}_C \otimes \mathcal{H}_{\mathcal{M}_v})$ is a time-dependent, unitary operator defined on the joint space of the coin and data qubits in node $v$, and $\mathds{1}_{\mathcal{M}_u}$ is the identity operator on the space of the data qubits in node $u$. The operator in \eqref{eq:local_op} is very general in that, if nodes implement arbitrary operators of this form along with shift operators, they implement the quantum walk control protocol in its full generality. It is possible to verify that coins satisfying \eqref{eq:extended_coin} have the form in \eqref{eq:local_op}.
In addition, \eqref{eq:local_op} directly extends to the case of multiple walks under the assumption that each walk interacts with a unique set of qubits in $\mathcal{M}$, or that walk interactions commute.

We now prescribe two operators following \eqref{eq:local_op} that are sufficient for universal distributed quantum computing and other network services, such as entanglement distribution, mediated by quantum walks.
The first generalizes the coin operator in \eqref{eq:extended_coin} to
\begin{align}
    & C(t) = \sum_{v} \dyad*{v} \otimes C_v(t) \otimes U_{v}(t) \label{eq:coin_op_assited},
\end{align}
where $U_v(t): \mathcal{H}_\mathcal{M} \to \mathcal{H}_\mathcal{M}$ is a unitary operator on the data space of the network written as
\begin{align}
    & U_v(t) = K_v(t) \bigotimes_{u \neq v} \mathds{1}_{\mathcal{M}_u},\label{eq:unit_expanded}
\end{align}
where $K_v(t): \mathcal{M}_v \to \mathcal{M}_v$ is a unitary operator on the space spanned by the data qubits in node $v$.
Essentially, $C(t)$ applies the operator $K_v(t)$ on qubits in $v$ if and only if the walker has a non-zero wavefunction component in $v$.
The operator $C(t)$ in \eqref{eq:coin_op_assited} extends the definition in \eqref{eq:extended_coin} to perform unitary operations on the data qubit space controlled by vertex position.
It allows for operations on the coin space of the walk through operators $C_v(t)$ and data-control interactions through operators $U_v(t)$, for every $v \in V$.
Consider the case where a set of qubits $\mathcal{Q}_v \subseteq \mathcal{M}_v$ in node $v$ is used to control unitary operations in the coin space of the walker. Let
\begin{align}
    & U_{\mathcal{Q}_v}(t) = \sum_{s \in \{0, 1\}^{|\mathcal{Q}_v|}} U_{vs}(t) \otimes \dyad{s} \bigotimes_{q' \notin \mathcal{Q}_v} \mathds{1}_{q'} \label{eq:qubit_control},
\end{align}
be a unitary operator acting on the coin space of vertex $v$ controlled by the qubits in $\mathcal{Q}_v$, defined on the joint space $\mathcal{H}_\mathcal{C} \otimes \mathcal{H}_\mathcal{M}$. The second operator has the form
\begin{align}
    & O(t) = \sum_{v} \dyad{v} \otimes U_{\mathcal{Q}_v}(t). \label{eq:data_control}
\end{align}

\subsection{Control-Control Interactions}

The QWCP allows for interactions between multiple quantum walks.
This is useful for expanding the control plane beyond distributed quantum computing, and brings flexibility to the implementation of distributed controlled quantum gates.
Interactions between quantum walks take the form of \eqref{eq:walk_interaction}, extended to $\mathcal{H}_g$ with $\mathds{1}_{\mathcal{M}}$, and do not change the state of data qubits in the network.
We focus on interaction operators for two quantum walks of the form
\begin{align}
    I_{vc}(t) = [(\mathds{1}_{\mathcal{H}_V^{2}} - \dyad*{v,v}) \otimes \mathds{1}_{\mathcal{H}_{\mathcal{C}}^{2}}] + \{\dyad*{v, v} \otimes [\dyad*c \otimes U_{vc} + (\mathds{1}_{\mathcal{H}_\mathcal{C}} - \dyad*{c}) \otimes \mathds{1}_{\mathcal{H}_\mathcal{C}}]\}, \label{eq:two_walk_interaction}
\end{align}
where $v \in V$ is an arbitrary node.
Note that we changed the order of the Hilbert space representation for two quantum walks from $\hilb_V \otimes \hilb_\mathcal{C} \otimes \hilb_V \otimes \hilb_\mathcal{C}$ to $\hilb_V^{2} \otimes \hilb_\mathcal{C}^{2}$ in order to simplify the description of $I_{vc}(t)$.
$I_{vc}(t)$ performs the unitary operation $U_{vc}$ in the coin space of the second walk controlled by the state $\ket{c}$ of the coin space of the first.
Operators of the form in \eqref{eq:two_walk_interaction} allow for the preparation of walk states that have Bell-state-like entanglement structures.
For instance, assume that the state of two walks at instant $t$ is
\begin{align}
    \ket{\Psi(t)}_W = \frac{\ket{v, c_0} + \ket{v, c_1}}{\sqrt{2}} \otimes \ket{v, c_0}.
\end{align}
By letting $U_{vc}$ as a permutation operator, the state
\begin{equation}
    \ket{\Psi(t)}_W = \frac{\ket{v, c_0, v, c_0} + \ket{v, c_1, v, c_1}}{\sqrt{2}}
\end{equation}
is obtained with one application of \eqref{eq:two_walk_interaction}.
GHZ-like entanglement for multiple quantum walks is similarly obtained by repeating the application of the control-control operation targeting different pairs of quantum walks at every application.

\section{Distributed quantum gates with one quantum walk}\label{sec:propagation}

The operators described in Section~\ref{sec:walk} constitute the fundamental operations of the QWCP and can be used to execute remote gates in a quantum network.
In this section, we apply the QWCP to implement distributed multi-qubit quantum gates with one quantum walk.
In particular, we define operations that allow a quantum walk to propagate control signals across paths of the network.
The results presented therein show that the QWCP is universal for distributed quantum computing.

\subsection{Path propagation and universality}\label{sec:universality}

Consider performing an arbitrary two-qubit controlled gate $CU: \hilb^{4} \to \hilb^{4}$ of the form
\begin{align}
    CU = \dyad*{0} \otimes \mathds{1}_{\hilb^{2}} + \dyad*{1} \otimes U, \label{eq:controlled_gate}
\end{align}
where $U: \hilb^{2} \to \hilb^{2}$ is a single-qubit gate, using qubit $a \in \mathcal{M}_{A}$ in node $A \in V$ as control and qubit $b \in \mathcal{M}_B$ in node $B \in V$ as target.
We now show how the protocol can be used to implement $CU$ and, thus, that it is universal for distributed quantum computing.
This last claim stems from the fact that a remote \ac{CNOT} gate can be executed between any two nodes of the network using the techniques presented therein.
To simplify notation, we only define terms for the operators $C(t)$ in \eqref{eq:coin_op_assited} and $O(t)$ in \eqref{eq:data_control} for the subspaces spanned by the qubits in nodes $A$ and $B$, considering undefined operators to be identities, and omit the qubits in $\mathcal{M} \setminus \{a, b\}$.
We use subscripts to refer to qubits, \textit{e.g} $\ket*{A, c_{Au}, 0_a, 1_b} \in \mathcal{H}_{W} \otimes \mathcal{H}^{4}$ represents the walker in edge $(A, u)$, and qubits $a$ in the $\ket*{0}$ state and $b$ in the $\ket*{1}$ state, where $\mathcal{H}^{4}$ is the Hilbert space spanned by 2 qubits.
Moreover, coin-space permutations are instrumental for the methods presented in this section.
Given two degrees of freedom $c_{vu}, c_{vw} \in \mathcal{C}_v$ for vertex $v \in V$, the operator of the form
\begin{align}
    & C_v^{c_{vu}c_{vw}} = \dyad{c_{vu}}{c_{vw}} + \dyad{c_{vu}}{c_{vw}} + \sum_{\substack{c \in \mathcal{C}_v \\ c \neq c_{vu}, c_{vw}}} \dyad{c} \label{eq:inner_coin}
\end{align}
permutes $c_{vu}$ with $c_{vw}$ while leaving all other degrees of freedom of $v$ unchanged.

We consider the initial state of the walker to be $\ket*{A, c_{A}}$, which corresponds to a self-loop in node $A \in V$. Thus, the global system is described by the state vector
$\ket*{\Psi(0)} = \ket*{A, c_{A}} \otimes (\alpha \ket*{0_a} + \beta \ket*{1_a}) \otimes \ket*{\Psi_b}$, \label{eq:separable}
where $\ket*{\Psi_b}$ is the state of $b$.
The interaction operator $O(t) = \sum_{v} \dyad{v} \otimes U_{\mathcal{Q}_v}(t)$ described in \eqref{eq:data_control} is applied with $U_{\mathcal{Q}_v} = U_{\{a\}}$ given by the controlled coin space permutation operation in \eqref{eq:inner_coin} as
\begin{align}
    & U_{\{a\}} = I_{\mathcal{C}_A} \otimes \dyad{0} + C_{A}^{c_{A}c} \otimes \dyad{1}, \label{eq:data_controlled_cnot}
\end{align}
where $c \neq C_A$ is any degree of freedom of $A$ and which generates the entangled state
\begin{align}
    & O(0) \ket*{\Psi(0)} = (\alpha \ket*{A, c_{A}, 0} + \beta \ket*{A, c, 1}) \otimes \ket*{\Psi_b}. \label{eq:entanglement_operator}
\end{align}
Our goal is to determine quantum walk operators that evolve this entangled state between control and data to a state of the form
$\ket*{\Psi(t)} = (\alpha \ket*{A,c_{A}, 0} + \beta \ket*{B, c_{B}, 1}) \otimes \ket*{\Psi_b}$.
When such a state is obtained, $CU$ is implemented with the application of the extended coin operator defined in \eqref{eq:coin_op_assited} and \eqref{eq:unit_expanded}, with $K_b = U$ and all other operators defined as identities. The final state obtained is
\begin{align}
    & C(t)\ket*{\Psi(t)} = \alpha \ket*{A, c_A, 0, \Psi_b} + \beta  (\mathds{1} \otimes U) \ket*{B, c_B, 1, \Psi_b} \label{eq:controlledU_state},
\end{align}
which shows the application of $CU$ controlled by $a$ with $b$ as a target.
Note that the state obtained in \eqref{eq:controlledU_state} is entangled with the walker subsystem; we will later demonstrate how to separate the data qubits from the walker.

\begin{figure*}
    \begin{subfigure}[b]{0.49\textwidth}
        \centering
        \includegraphics[scale=0.2]{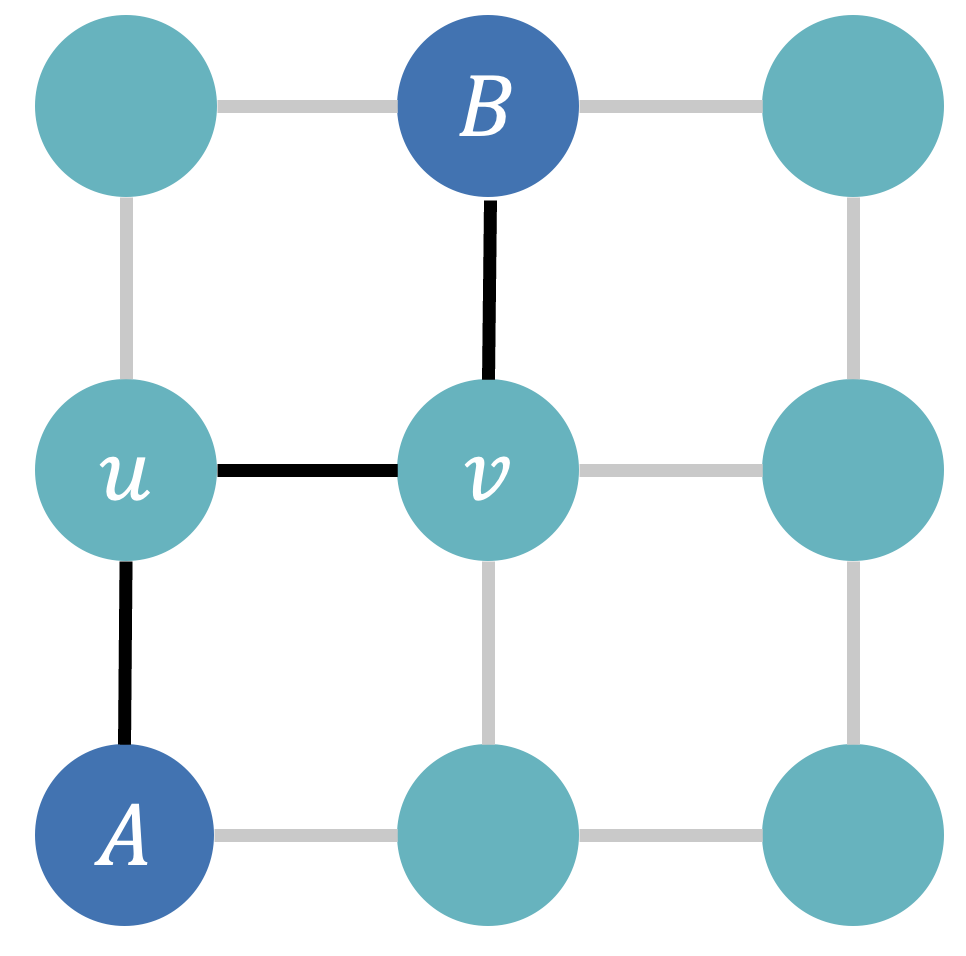}
        \caption{Path $p$ connecting $A$ and $B$.}
        \label{subfig:path_selection}
    \end{subfigure}
    \begin{subfigure}[b]{0.49\textwidth}
        \centering
        \includegraphics[scale=0.15]{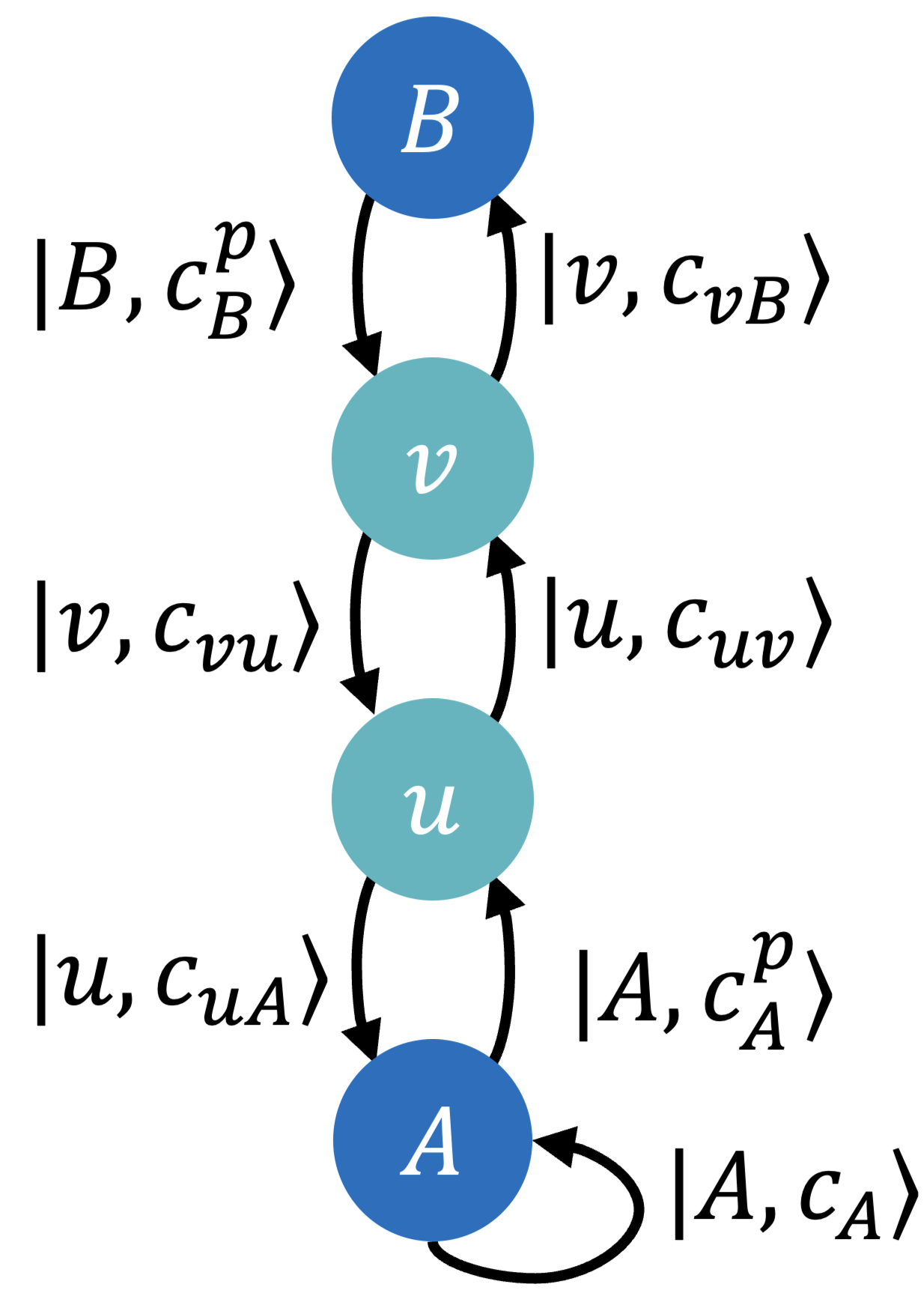}
        \caption{Map between edges and vectors.}
        \label{subfig:path_chiralities}
    \end{subfigure}
    \caption{Notation for edges exemplified in a grid graph. Consider that $A$ and $B$ are two nodes connected in a 2D grid network. (\subref{subfig:path_selection}) $p$ is a minimum path connecting $A$ and $B$ with hop-distance $3$ traversed by the walker. (\subref{subfig:path_chiralities}) Each edge on the path corresponds to a vector in $\mathcal{H}_W$, which appear in the walker wavefunction throughout movement. The degrees of freedom are defined such that $\ket*{x, c_{xy}}$ represents edge $(x, y)$. As an example, the flip-flop operator specified in \eqref{eq:flipflop} maps $\ket*{v, c_{vu}} \to \ket*{u, c_{uv}}$, while the operator $C_u$ defined in terms of  \eqref{eq:inner_coin} maps $\ket*{u, c_{uA}} \to \ket*{u, c_{uv}}$.}
    \label{fig:routing_grid}
\end{figure*}

Quantum walk evolution is restricted to neighbor locality, such that, to have the state given in \eqref{eq:controlledU_state} at time $t$, all of the wavefunction at time $t - 1$ must exclusively be a superposition of edges incident to nodes $A$ and $B$, and its neighbors. There are many ways to define coin and shift operators with this behavior and we focus on the case where the quantum walk traverses a single path connecting $A$ and $B$. Some auxiliary definitions and assumptions are required to describe the operators in this context. Let $p$ be a path of the network connecting $A$ and $B$ with hop distance $\Delta_p(A, B)$. We assume every node knows the network topology and that classical information can be transmitted between the nodes. Recall that the edges of $N'$  are mapped to walker states following the relation
$(v, u) \to \ket*{v, {c_{vu}}}$,
and that the self-loop $(v, v)$ is mapped to the degree of freedom $c_v$, for all $v$. We refer to the edge that connects $A$ to its neighbor in $p$ as $\ket*{A, c_{A}^{p}}$ and the reverse edge that connects $B$ to its preceding vertex in $p$ as $\ket*{B, c_{B}^{p}}$. This notation is depicted in Figure \ref{fig:routing_grid} for a 2D-grid network. For simplicity, assume that propagation starts at time $t = 0$. Neighbor locality implies that $\Delta_p(A, B)$ steps are required to complete walker propagation through $p$.

\begin{figure}
\begin{centering}
\begin{subfigure}[b]{0.23\textwidth}
    \centering
    \includegraphics[scale=0.15]{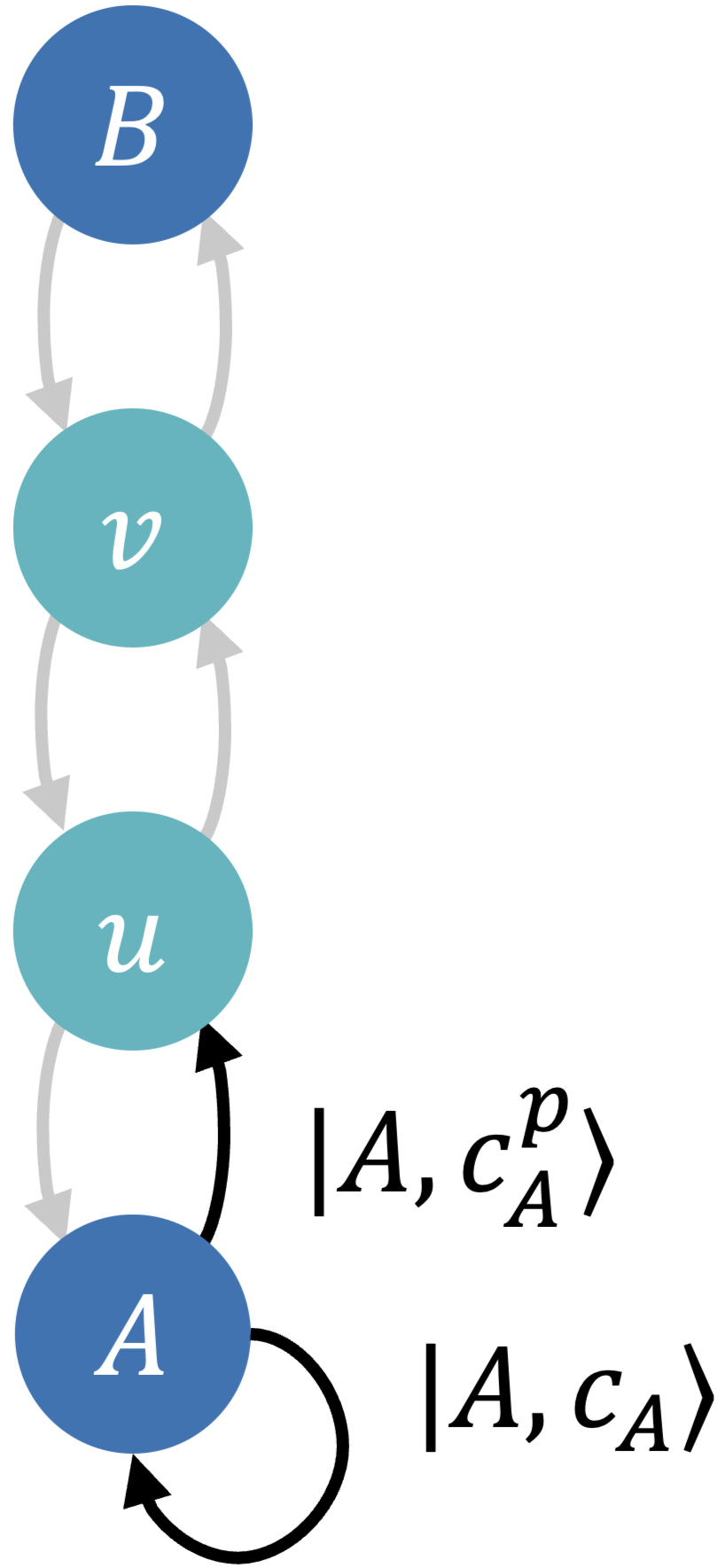}
    \caption{Coin in $A$.}
    \label{subfig:path1}
\end{subfigure} \hfill
\begin{subfigure}[b]{0.23\textwidth}
    \centering
    \includegraphics[scale=0.15]{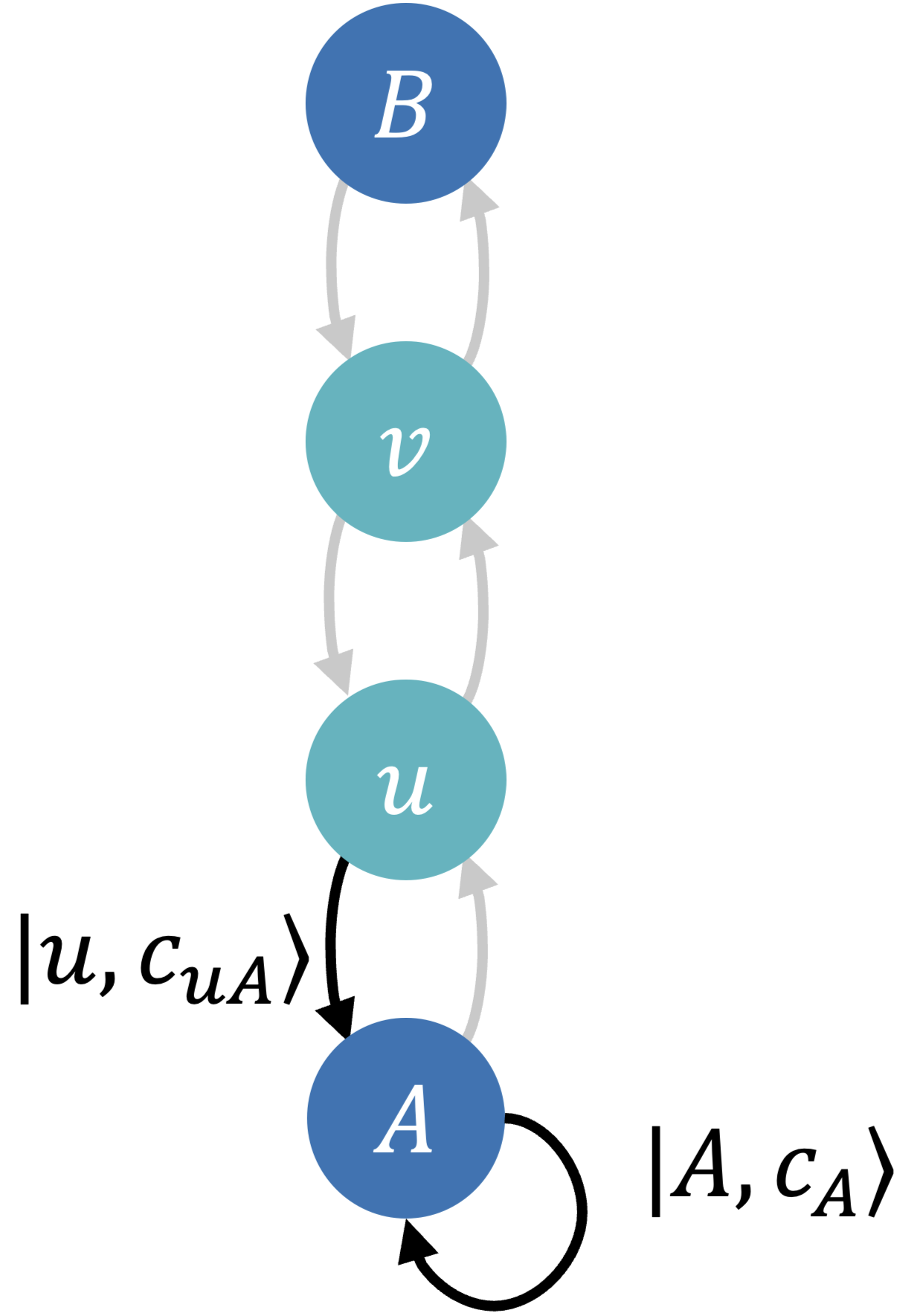}
    \caption{Flip-flop shift}
    \label{subfig:path2}
\end{subfigure} \hfill
\begin{subfigure}[b]{0.23\textwidth}
    \centering
    \includegraphics[scale=0.15]{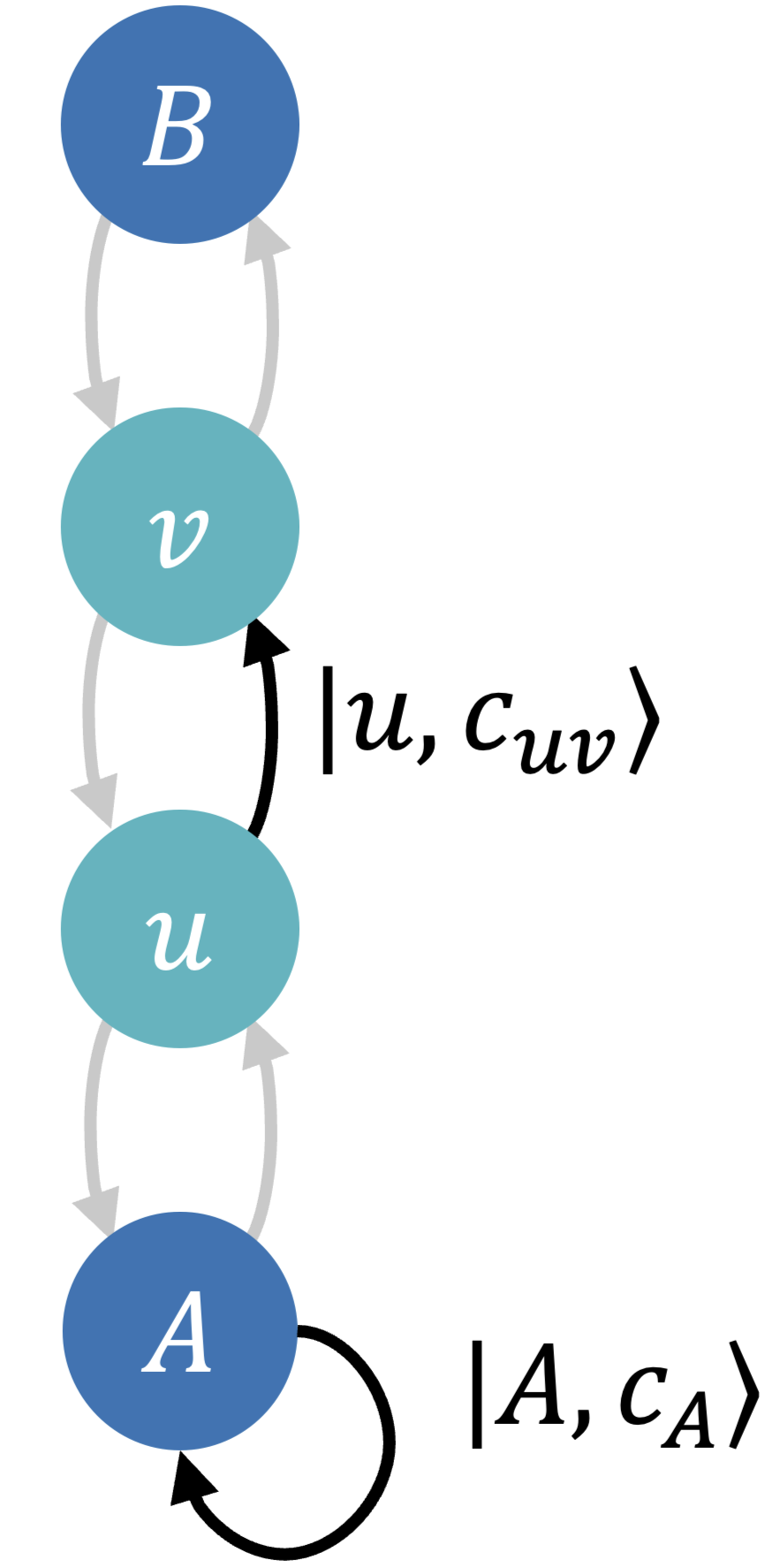}
    \caption{Coins in path.}
    \label{subfig:path3}
\end{subfigure}
\begin{subfigure}[b]{0.23\textwidth}
    \centering
    \includegraphics[scale=0.15]{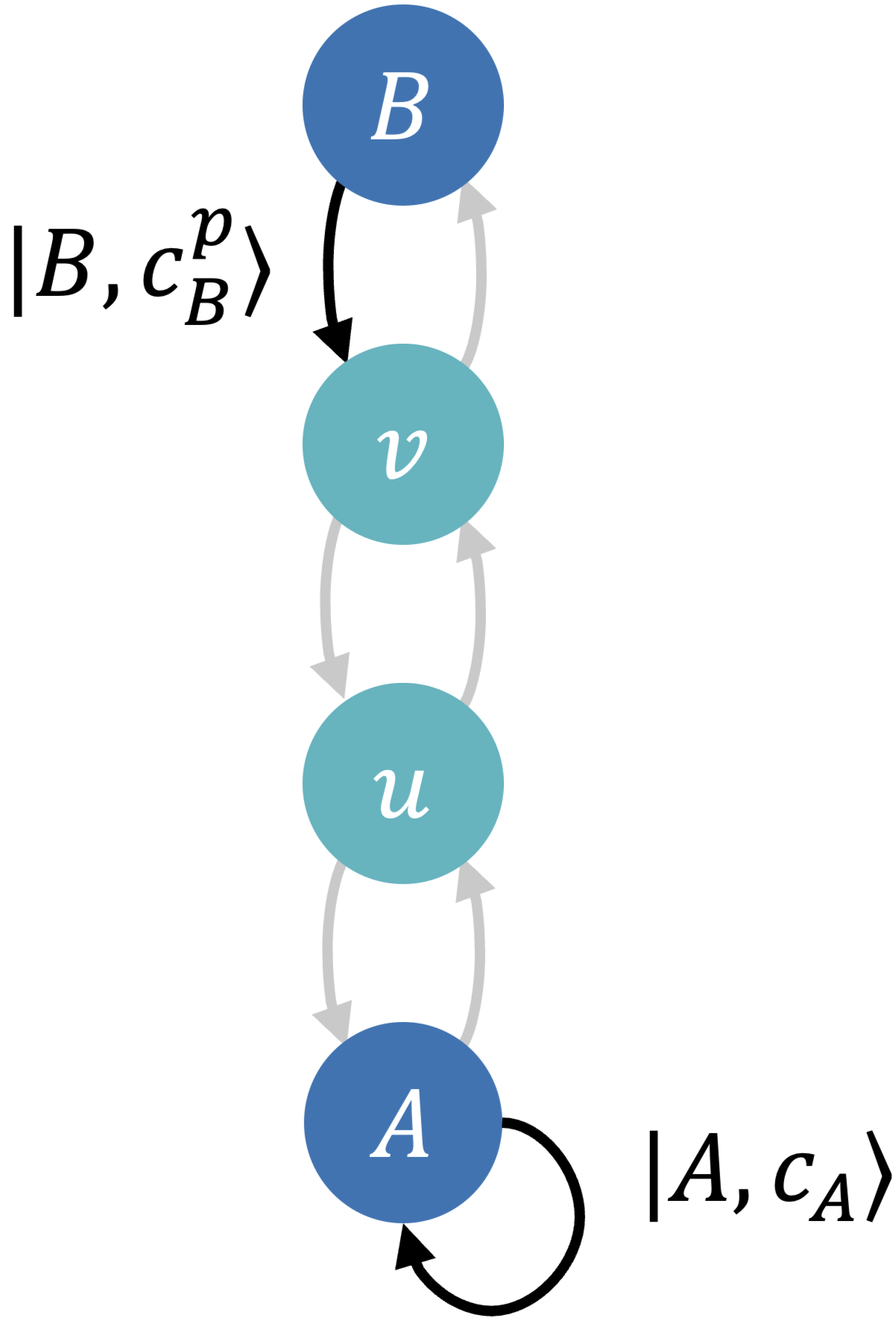}
    \caption{Final state.}
    \label{subfig:path4}
\end{subfigure}
\end{centering}
\caption{Protocol execution in a $3$-by-$3$ grid with a quantum walk through a path $p$. Dark edges depict vectors which have non-zero wavefunction component in a given step. (\subref{subfig:path1}) The initial state of execution is generated by the application of the controlled operation demonstrated in \eqref{eq:entanglement_operator} with the coin permutation operator in~\eqref{eq:inner_coin}. (\subref{subfig:path2}) The flip-flop shift exchanges edge $(A, u)$ with edge $(u, A)$, moving the walker while mapping the self-loop edge to itself. (\subref{subfig:path3}) After the first coin flip, all subsequent coin operators work as shift operators inside a node, mapping degrees of freedom in order to propel the walker towards $B$. (\subref{subfig:path4}) After $\Delta_p(A, B) = 3$ steps, the final wavefunction is a uniform superposition between edges $(A, A)$ and $(B, w)$, which can be used to perform an operation controlled by qubit $a$ located in $A$ with target qubit $b$ located in $B$.}
\label{fig:walk_assisted}
\end{figure}

Path propagation starts with the state in~\eqref{eq:entanglement_operator} with $c = C_{A}^{p}$, as depicted in Fig.\ref{subfig:path1}.
We use the extended flip-flop shift operator, \eqref{eq:extended_shift}, to route information during all time steps, a behavior portrayed in Fig.\ref{subfig:path2}.
The coin operator is also time-independent, although it depends on the path $p$ chosen.
All operators $C_v$ in \eqref{eq:coin_op_assited} follow \eqref{eq:inner_coin} and have the form
$C_{v}^{p} = C_{v}^{c_1c_2}$, where $c_1$ and $c_2$ refer to the degrees of freedom that represent the edges incident to $v$ in $p$.
Thus, we define $C_v^{p}$ specifying $c_1$ and $c_2$ for the vertices of interest, and assume that $C_{u}(t) = \mathds{1}_{\mathcal{C}_u}$ for all nodes $u \notin p$ and for node $A$.
Let $w$ and $v$ be the neighbors of $u \in p \setminus\{A, B\}$ on the path $p$.
$C_u$ has $c_1 = c_{uw}$ and $c_2 = c_{uv}$, representing a permutation between edges $(u, w)$ and $(u, v)$ in $p$ that is shown in Fig.\ref{subfig:path3} for $w = A$.
The operator $C_B^{p}$ has $c_1 = c_B^{p}$ and $c_2 = c_{B}$, providing the permutation behavior depicted in Fig.\ref{subfig:path4}.
It suffices to set $K_v(t) = I_{\mathcal{M}_v}$ in \eqref{eq:unit_expanded} to perform the desired controlled operation between $a$ and $b$, although it is possible to perform operations controlled by $a$ on the qubits in the intermediate nodes as the walker moves by choosing $K_v(t)$ accordingly.
In particular, assume that gate $U_v$ controlled by qubit $a$ in $A$ must be performed on the data qubits in $v$, for all $v \in p \setminus\{A, B\}$.
The gates are implemented by taking $K_v(\Delta_p(A, v)) = U_v$ for all $v \in p$.

Quantum control through path propagation is directly extended to gates controlled by multiple qubits in $A$, such as a Toffoli gate, and for gates with multiple targets.
Toffoli-like behavior is achieved by using initial interaction operators of the form
\begin{align}
    U_{\mathcal{Q}_A} = I_{\mathcal{C}_A} \otimes (\mathds{1} - \dyad{s}) + C_{A}^{c_{A}c} \otimes \dyad{s}, \label{eq:gen_data_controlled}
\end{align}
where $s$ is a $|\mathcal{Q}_A|$-bit string that determines the state of the data qubits $\mathcal{Q}_A$ in node $A$ used for control.
For gates with multiple targets, it suffices to modify the operator $K_v(t)$ accordingly.


The overall behavior of the walker is illustrated in Figure \ref{fig:walk_assisted}.
The initial state in \eqref{eq:entanglement_operator} is a superposition between the states $\ket*{A, c_{A}}$ and $\ket*{A, c_{A}^{p}}$ that are entangled with qubit $a$ in node $A$.
The flip-flop shift propagates the state component $\ket*{A, c_{A}^{p}}$ along $p$ and ensures that $\ket*{A, c_{A}}$ remains in superposition throughout protocol execution.
Coin operators act as forwarding operations in the nodes ensuring propagation.
The net effect of $\Delta_p(A, B)$ successive applications of $S(t)C(t)$ is the superposition specified in \eqref{eq:controlledU_state}.

\subsection{Separating data and control} \label{subsec:separation}

The entangled state prescribed in \eqref{eq:controlledU_state} includes both control and data qubits.
Hence a partial trace operation on the walker system will not leave the state of $a$ and $b$ as it would be if the operation where performed without the walker.
We propose two solutions in order to overcome this problem both of which can be generalized to the case of more complex quantum gates distributed by multiple quantum walks.

The first technique consists of control operations directed at concentrating the walker's wavefunction into a single network node, such as by propagating the walker wavefunction from both $A$ and $B$ to an intermediate node.
This operations can be performed in parallel with any further quantum operations that nodes $A$ and $B$ may perform on their data qubits and does not impact computation time.
For simplicity, we report a strategy where the walker evolution is reversed once the coin operation that performs $CU$ takes place by applying inverses of the unitary path propagation operators.
Since the propagation operators are permutation operators, they are Hermitian unitaries and, thus, are their own inverses.
It takes $\Delta_p(A,B)$ time steps to reverse the walker back to $A$ and transform the joint state of the system to the form
$\alpha \ket*{A, c_{A}, 0_a, \Psi_b} + \beta (\mathds{1} \otimes U) \ket*{A, c, 1_a, \Psi_b}$.
To finish separation, the data controlled CNOT operation specified in \eqref{eq:data_controlled_cnot} provides the state 
$\ket*{A, c_{A}} \otimes (\alpha \ket*{0_a, \Psi_b} + \beta (\mathds{1} \otimes U) \ket*{1_a, \Psi_b})$,
which is a separable state between control and data.
The separability between the walker system and qubits $a$ and $b$ ensures that the state of the qubits after a partial trace operation on the walker system is the state that one obtains if $a$ and $b$ were in the same node and a $CU$ gate was applied locally.

The second method relies on local measurements of control registers in nodes $A$ and $B$ that do not destroy the state of qubits $a$ and $b$.
For ease of explanation, assume $C_A = C_B = 0$, which can always be performed with a coin permutation operator.
Moreover, let $\mathcal{B}_{AB}$ be a basis for the vertex space $\mathcal{H}_V$ that contains the orthogonal vectors
\begin{equation}
    \ket*{AB_{+}} = \frac{\ket*{A} + \ket*{B}}{\sqrt{2}} \text{ and } \\
    \ket*{AB_{-}} = \frac{\ket*{A} - \ket*{B}}{\sqrt{2}}.
\end{equation}
Let $\ket*{\Psi_{ab}^{\pm}} = \alpha \ket*{0, \Psi_b} \pm \beta  (\mathds{1} \otimes U) \ket*{1, \Psi_b}$.
In the basis $\mathcal{B}_{AB}$ for the vertex space, the entangled state in \eqref{eq:controlledU_state} becomes
\begin{align}
    \frac{\ket*{AB_{+}} \ket*{\Psi_{ab}^{+}} + \ket*{AB_{-}} \ket*{\Psi_{ab}^{-}}}{\sqrt{2}}.
\end{align}
The structure of this last state shows that a projective measurement of the vertex space in the basis $\mathcal{B}_{AB}$ will decouple control and data, leaving the data qubits in the desired state up to a $Z$ gate correction in qubit $a$ that depends on the classical values of the measurement outcome.
We now demonstrate that measurements in $\mathcal{B}_{AB}$ are indeed local operations and represent local $X$ basis measurements of the qubits spanning the vertex space.
$A = A_1 \ldots A_{k}$ and $B = B_1 \ldots B_{k}$ are two $k$-bit binary strings, where $k$ depends on a particular implementation of the quantum walk.
Naturally, $j \geq 1$ bits must differ in $A$ and $B$, otherwise $A = B$.
Without loss of generality, assume that the distinct bits are the last $j$ bits in $A$ and $B$, i.e., $A = A_{1} \ldots A_{k}$ and $B = A_{1} \ldots A_{k - j} \overline{A}_{k - j + 1} \ldots \overline{A}_{k}$ and the basis states can be written as
\begin{equation}
    \ket*{AB_{\pm}} = \ket*{A_{1} \ldots A_{k - j}} \otimes \frac{\ket*{A_{k - j + 1} \ldots A_{k}} \pm \ket*{\overline{A}_{k - j + 1} \ldots \overline{A}_{k}}}{\sqrt{2}},
\end{equation}
where the state after the tensor product sign is a generalized $j$-qubit GHZ state.
It is known from the structure of GHZ states that local measurements in the $X$ basis yield the desired result.
In particular, a projective measurement in a basis following $\mathcal{B}_{AB}$ can be performed by locally measuring the first $k - j$ vertex space qubits in the $Z$ basis and the remaining $j$ qubits in the $X$ basis.
The distributed implementation of the quantum walk implies that, in general, the vertex qubits are spread across multiple nodes in the network.
Nonetheless, the measurements considered are local and can be implemented without the need for distributed entanglement across the nodes.
Finally, the measurement outcomes must be transmitted to node $B$ via classical communication to control a correction operation.

Separating control and data with the methods described imposes different requirements on utilization of computing and networking resources.
Reversing the quantum walk evolution does not increase computation time as operations on the data qubits can be executed concurrently, although it demands the use of quantum channels for separability after a control operation is performed.
On the other hand, local measurements only require the exchange of classical communication among nodes, albeit they increase computation time when quantum circuits performed in node $B$ do not commute with $Z$ gates.
Therefore, the methods can be used to optimize resource utilization based on different goals in order to improve the efficiency of the QWCP under distinct scenarios.

\subsection{Controlling operations with qubits in different nodes}

The path propagation procedure is easily modified to allow for execution of remote gates controlled by multiple qubits at different nodes.
For instance, consider the case of a Toffoli gate targeting qubit $b$ in node $B$ controlled by qubits $a_0$ and $a_1$ in nodes $A_0$ and $A_1$, respectively.
Path propagation can be used to implement the gate by sending the walker from $A_0$ to $B$ through a path passing through node $A_1$.
Following the results presented for path propagation, it suffices to demonstrate how the quantum walk state is capable of acquiring the proper dependence with the state of qubits $a_0$ and $a_1$.
We now show how a Toffoli gate is performed, focusing on the joint state of the quantum walk system with qubits $a_0$ and $a_1$.
The techniques are directly applicable to gates with arbitrary numbers of control qubits at different nodes.

Assume that the quantum walk is initialized at node $A_0$.
The initial procedure used for path propagation, which applies the interaction operator in \eqref{eq:data_control} using the coin permutation operator in \eqref{eq:inner_coin} generates a state of the form
\begin{align}
    & \ket*{\Psi(0)} = (\alpha_0 \ket*{A_0, c_{A_0}, 0_{a_0}} + \beta_0 \ket*{A_0, c_0, 1_{a_0}}) \otimes (\alpha_1 \ket*{0_{a_1}} + \beta_1 \ket*{1_{a_1}}), \label{eq:toffoli_entangled}
\end{align}
where $\alpha_j, \beta_j$ are the amplitudes defining the state of control qubit $a_j$, for $j \in \{0, 1\}$, and $c_0 \in \mathcal{C}_{A_0}$ is a given degree of freedom of $A_0$.
The state in \eqref{eq:toffoli_entangled} is similar to that in \eqref{eq:entanglement_operator}, although it shows the states of the control qubits explicitly.
Once this initial state is prepared the walk from $A_0$ across path $p$ to $A_1$ yields a state of the form
\begin{align}
    & \ket*{\Psi(\Delta_p(A_0, A_1))} = (\alpha_0 \ket*{A_{0}, c_{A_0}, 0_{a_0}} + \beta_0 \ket*{A_1, c_{A_1}, 1_{a_0}}) \otimes (\alpha_1 \ket*{0_{a_1}} + \beta_1 \ket*{1_{a_1}}). \label{eq:toffoli_entangled_2}
\end{align}
An interaction operator between data and control using the controlled coin permutation operator $C_{A_1}^{c_{A_1}c_1}$ following \eqref{eq:inner_coin}, where $c_1 \in \mathcal{C}_{A_1}$, yields
\begin{align}
    \alpha_0\alpha_1 \ket*{A_{0}, c_{A_0}, 0_{a_0}, 0_{a_1}} + \alpha_0\beta_1 \ket*{A_{0}, c_{A_0}, 0_{a_0}, 1_{a_1}} + \beta_0 \alpha_1 \ket*{A_1, c_{A_1}, 1_{a_0}, 0_{a_1}} + \beta_0 \beta_1 \ket*{A_1, c_1, 1_{a_0}, 1_{a_1}}. \label{eq:toffoli_control}
\end{align}
The component $\ket*{A_1, c_1}$ of the state carries the necessary control for a the Toffoli gate controlled by $a_0$ and $a_1$, and can be transmitted to node $B$.
In this case, the Toffoli gate can be implemented at time $\Delta_p(A_0, A_1) + \Delta_p(A_1, B)$ using a coin operator following \eqref{eq:coin_op_assited} and \eqref{eq:unit_expanded}, with $K_B$ performing an $X$ gate on qubit $b$.
Generic $n$-qubit gates, which are controlled operations of the form
\begin{align}
    CU = \dyad*{s} \otimes U + (\mathds{1} - \dyad*{s}) \otimes \mathds{1}
\end{align}
for some $n'$-bit string $s$, with $n' < n$ and $U: \hilb^{2^{(n - n')}} \to \hilb^{2^{(n - n')}}$ can be implemented using this procedure.

The single-path propagation procedure described in Section~\ref{sec:universality} suffices for universal quantum computing.
This implies that the aforementioned multi-qubit controlled gates can be implemented by first decomposing the gate into two-qubit gates and then performing each of these gates with the control protocol.
The results shown in this section add to the initial path-propagation scheme by providing a way to implement these gates using network resource differently.

\section{Parallel Control Propagation with Multiple Quantum Walks}\label{sec:parallel}

We applied the QWCP in the previous section to implement distributed quantum gates.
This extends to parallel control signal propagation with multiple quantum walks, as we show next.
Let $A$ denote the node in the network containing control qubits.
Let $B^{k} =\{B_0, \ldots, B_{k - 1}\}$ denote a set of target nodes.
For simplicity, assume the goal of performing gate $CU_j: \hilb^{4} \to \hilb^{4}$ following~\eqref{eq:controlled_gate} controlled by qubit $a$ in node $A$ with target qubit $b_j$ in node $B_j$, for $j = 0, \ldots, k - 1$.
There are two extensions to propagate control signals in order to perform these gates in parallel.
The first allows multiple entangled quantum walks through distinct paths.
The second propagates entangled control signals through a directed rooted tree in the network.

\subsection{Control fan-out signals}

We introduce a branching operator for quantum walks in order to present parallel propagation with entangled walkers.
This operation consists of successive applications of the control-control interaction operator, \eqref{eq:two_walk_interaction}, to generate entangled quantum walks as follows.
Suppose a quantum walker is received at time $t$ at node $v \in V$ and $k$ walkers entangled with the received walker must each be propagated to a neighbor $v_j \in \delta(v)$ of $v$, for $j \in \{0, \ldots, k - 1\}$.
Let $\mathcal{V}_v = \{v_0,\ldots, v_{k - 1}\} \subseteq \delta(v)$ denote the set of neighbors of $v$ to which these $k$ control signals must be propagated.
Without loss of generality, assume the state of the walker received at node $v$ is
$\ket*{\Psi_W} = \alpha \ket*{A, c_A} + \beta \ket*{v, c}$,
where $\alpha, \beta \in \mathbb{C}$ and $c \in \mathcal{C}_{v}$ is any degree of freedom of the walker.
$k - 1$ walkers are initialized at node $v$, and the joint state of all $k$ quantum walks is 
\begin{align}
    & \ket*{\Psi_W(t)} = (\alpha \ket*{A, c_A} + \beta \ket*{v, c_v}) \otimes \ket*{v, c_v}^{\otimes k - 1}.
\end{align}
Applying operator $I_{vc_{v}}$ given by \eqref{eq:two_walk_interaction} with $U_{vc} = C_{v}^{c c_{vv_j}}$ following \eqref{eq:inner_coin}, using the first walk as control and the $j$-th walk as target, for each $v_j \in \mathcal{V}_v$, yields the GHZ-like entangled state of form
\begin{align}
    & I_{\mathcal{V}_v} \ket*{\Psi(t)} = \alpha (\ket*{A, c_A} \otimes \ket*{v, c_v}^{\otimes k - 1}) + \beta \ket*{v, c, v, c_{vv_1},\ldots, v, c_{vv_{k - 1}}},
\end{align}
where $I_{\mathcal{V}_v}$ denotes the product of all control-control operators $I_{vc_{v}}$ used.
The 1-to-$k$ control fan-out operator $F_{c\mathcal{V}_v}: \hilb_{W}^{k} \to \hilb_{W}^{k}$ has the form
\begin{align}
    & F_{c\mathcal{V}_v} = (C_{c\mathcal{V}_v} \otimes \mathds{1}_{\hilb_{W}^{k - 1}}) I_{\mathcal{V}_v},\label{eq:fanout}
\end{align}
where $C_{c\mathcal{V}_v}: \hilb_{W} \to \hilb_{W}$ is the coin operator for the first quantum walk given by
\begin{align}
    & C_{c\mathcal{V}_v} = (\dyad*{v} \otimes C_{v}^{c c_{vu_0}}) +  (\mathds{1}_{\hilb_V} - \dyad*{v}) \otimes \mathds{1}_{\hilb_c},
\end{align}
with $C_v^{c c_{vu_0}}$ following~\eqref{eq:inner_coin}.
$F_{c\mathcal{V}_v}$ yields an entangled state of the $k$ walkers with a component $\ket*{v, c_{vu_0}, \ldots v, c_{vu_{k - 1}}}$ where the $j$-th walker points to the edge $(v, u_j)$ for $j = 0,\ldots, k - 1$.
The shift operator moves each walker in superposition to the corresponding neighbor of $v$.
Note that the first quantum walker received at $v$ is sent to node $u_0$ and 1-to-$k$ control fan-out initializes $k - 1$ walkers in $v$.

\subsection{Parallel control through multiple paths}\label{sec:multi-path}

Let $\mathcal{P} = \{p_0, \ldots, p_{k - 1}\}$ denote a set of paths where $p_{j}$ starts at $A$ and ends at $B_{j} \in B^{k}$ for $j = 0,\ldots, k - 1$. 
The analysis of a single walker system extends to multiple walker systems by using one quantum walk to propagate control signals through each path in $\mathcal{P}$.
The process starts with the initialization of one walker in $A$ and the application of interaction operator in \eqref{eq:data_control} to generate the entangled state between walker and data shown in~\eqref{eq:entanglement_operator}, where $c_A^{p} = c_A^{p_0}$ in this case.
Let $\mathcal{V}_A^{\mathcal{P}}$ denote the set of neighbors of $A$ in each path in $\mathcal{P}$.
The control fan-out operator $F_{c_A^{p_0}\mathcal{V}_A^{\mathcal{P}}}$, ~\eqref{eq:fanout}, is applied to prepare one walker for each path in $\mathcal{P}$.
Assign walker $j$ to path $p_j$.
Shift operators for all quantum walkers are time-independent and follow \eqref{eq:extended_shift}.
The coin operator $C^{p_j}(t): \hilb_W^{k} \to \hilb_W^{k}$ for walker $j$ is defined to obtain the propagation behavior depicted in Fig.\ref{fig:walk_assisted}.
The separability of the individual coins for each walker yields the $k$-walker coin operator $C^{\mathcal{P}}(t): \hilb_W^{k} \to \hilb_W^{k}$ for the set of paths $\mathcal{P}$ of the form
\begin{align}
    & C^{\mathcal{P}}(t) = \bigotimes_{j = 0}^{k - 1}C^{p_j}(t). \label{eq:sep_path_coin}
\end{align}
Walker $j$ reaches node $B_j$ after $\Delta(A, B_j)$ steps of evolution and gate $CU_j$ is implemented by applying the extended coin operator in~\eqref{eq:coin_op_assited}~and~\eqref{eq:unit_expanded} with $K_{B_j} = U_j$.



\subsection{Parallel control through trees}\label{sec:tree_propagation}

Suppose we are given a directed tree $\mathcal{T} = (V_\mathcal{T}, E_\mathcal{T}) \subseteq N$ rooted in node $A$ containing the target nodes in $B^{k}$.
Let $p(v) \in \delta(v)$ denote the predecessor of $v \in V_\mathcal{T} \setminus \{A\}$ in $\mathcal{T}$.
Let $\mathcal{S}(v) \subset \delta(v)$ denote the set of successors of node $v \in V_\mathcal{T}$ in $\mathcal{T}$.
We now define operations for the parallel propagation of control signals through $\mathcal{T}$ such that each node in $V_\mathcal{T} \setminus \{A\}$ receives exactly one walker throughout the process.
For simplicity, we define coin operators as specified in~\eqref{eq:coin_op_assited} by determining $C_v(t)$ for the nodes in $V_\mathcal{T}$, including the coins required for the fan-out operation in~\eqref{eq:fanout}, and define $C_u(t) = \mathds{1}_{\mathcal{M}_u}$ for $u \notin \mathcal{T}$.
Coins for nodes $B_j \in B^{k}$ have $U_v(t) = U_j \bigotimes_{u \neq v} \mathds{1}_{\mathcal{M}_u}$ in order to realize gates $CU_j$, while coins for nodes $v \in V_\mathcal{T} \setminus B^{k}$ have $U_v(t) = \mathds{1}_\mathcal{M}$ and reduce to the form in~\eqref{eq:extended_coin}.
Similar to the case for multiple paths, the process starts at $A$ with the preparation of an entangled state between data and control as shown in~\eqref{eq:entanglement_operator}, and all shift operators follow~\eqref{eq:extended_shift}.
There are three possible control operations performed at each node $v \in V_\mathcal{T}$.
If $|\mathcal{S}(v)| = 1$, $C_v(t) = C_{v}^{c_{1}c_{2}}$ following~\eqref{eq:inner_coin} with $c_1 = c_{vp(v)}$ and $c_2 = c_{vu}$, where $u \in \mathcal{S}(v)$ denotes the unique successor of $v$ in $\mathcal{T}$.
When $|\mathcal{S}(v)| > 1$, the control fan-out operator $F_{c \mathcal{V}_v}$ given by~\eqref{eq:fanout} is applied with $c = c_{vp(v)}$ and $\mathcal{V}_v = \mathcal{S}(v)$.
If $|\mathcal{S}(v)| = 0$, i.e., $v$ is a leaf of $\mathcal{T}$, $C_v(t) = \mathds{1}_{\mathcal{C}_v}$.
In the case of node $A$, $c_{v p(v)}$ is replaced with the degree of freedom $c$ used in~\eqref{eq:entanglement_operator}.
A walker system reaches node $v \in \mathcal{T}$ at time $t = \Delta_{\mathcal{T}}(A, v)$, where $\Delta_{\mathcal{T}}(A, v)$ is the hop distance between $A$ and $v$ in $\mathcal{T}$.

\section{Application to Entanglement Distribution}\label{sec:distribution}

In this section, we apply the QWCP to implement entanglement distribution in the network.
We also discuss how entanglement distribution protocols defined in the literature~\cite{chakraborty2020entanglement,li2021effective,meignant2019distributing, bugalho2023distributing,pant2019routing, patil2022entanglement} can be deployed using QWCP operators.

\subsection{Entanglement distribution through quantum gate distribution}

A straightforward way to perform entanglement distribution with the QWCP is to use the protocol to distribute local quantum circuits that generate desired entangled states.
The operations specified for distributing quantum gates in Sections~\ref{sec:propagation}~and~\ref{sec:parallel} are directly applicable in this case.
Each multi-qubit gate required in an entangled state preparation circuit can be distributed through quantum walks using one or multiple paths, or network trees.
For instance, an $n$-qubit GHZ state can be locally generated with one Hadamard and $n - 1$ CNOT gates, and each CNOT can be distributed using the QWCP.

\subsection{Generating GHZ states through paths and trees}

We now shift gears to detail how the parallel control propagation operations can be modified to implement GHZ-state distribution on paths and trees.
We describe the main idea for paths, and the extension for trees is obtained by replacing multi-path control propagation with tree control propagation.
Suppose that a set of $k$ paths $\mathcal{P} = \{p_0, \ldots, p_{k - 1}\}$ in the network is given.
The goal is to generate $k$ independent GHZ states across data qubits at each node of each path in $\mathcal{P}$.
Let $\mathcal{Q}_{jv}$ denote the set of data qubits at node $v$ in path $p_j$ which must be part of the same GHZ state, for $j \in \{0, \ldots, k - 1\}$, and $\mathcal{Q}_j = \bigcup_{v \in p_j} \mathcal{Q}_{jv}$.
For simplicity, let $X_{\mathcal{Q}_{jv}}$ denote the application of the single-qubit Pauli $X$ gate on all qubits in $\mathcal{Q}_{jv}$ extended to the entire data plane using the identity operator.
Since GHZ states related to different paths are independent, $\mathcal{Q}_{i} \cap \mathcal{Q}_{j} = \{\}$ for $i, j \in \{0,\ldots, k - 1\}$ and $i \neq j$.
Let $A_j$ and $B_j$ denote the first and last nodes in path $p_j$, respectively.
The process starts in parallel at all nodes $A_j$, $j \in \{0, \ldots, k - 1\}$, at time $t = 0$ by locally preparing the data qubits in $\mathcal{Q}_{jA}$ in a GHZ state. The data controlled interaction operator in~\eqref{eq:data_control} is applied to generate the following entangled state
\begin{align}
    & \ket{\Psi(0)} = \bigotimes_{j = 0}^{k}\frac{1}{\sqrt{2}} (\ket*{A_j, c_{A_j}}\ket*{00\ldots0}_{\mathcal{Q}_{jA_j}} + \ket*{A_j, c_{A_j^{p_j}}}\ket*{11\ldots1}_{\mathcal{Q}_{jA_j}}).\label{eq:initial_ghz_state}
\end{align}
The desired GHZ states are obtained by applying the coin and shift operations specified in Section~\ref{sec:propagation} for each path in $\mathcal{P}$, with a slight modification in the coin operators.
In this case, the behavior of the coin operation, \eqref{eq:coin_op_assited}, for the $j$-th walker at node $v \in p_j \setminus \{A_j\}$ is of the form $C_v(t) \otimes X_{\mathcal{Q}_{jv}}$.
Thus, at time step $\Delta_{p_j}(A, v)$, the coin operation expands the GHZ states of data qubits with qubits in $\mathcal{Q}_{jv}$.
The process for path $p_j$ terminates at time $t = \Delta_{p_j}(A, B_j)$, and the overall distributions ends at time $t = \max_j \Delta_{p_j}(A, B_j)$.

\subsection{Link-level entanglement generation with QWCP}

Parallel \textit{link-level entanglement generation} is a building block for entanglement distribution protocols~\cite{chakraborty2020entanglement, li2021effective, pant2019routing, patil2022entanglement}.
In the process, neighboring network nodes use the channels that interconnect them to generate maximally entangled states.
Link-level entanglement generation is performed with quantum walks as follows, where we omit the state of data qubits for simplicity.
Let $(u, v) \in E$ and assume, without loss of generality, that $u < v$.
For every pair $u, v$, a quantum walk in state $\ket{\Psi} = \ket{u, c_u}$ is initialized at node $u$, leading to $k = |E| / 2$ quantum walks initialized in $\mathcal{H}_\mathcal{N}$.
A coin operator $C_u$ following~\eqref{eq:extended_coin} that maps $\ket{u, c_u} \to (\ket*{u, c_u} + \ket*{u, c_{uv}}) / \sqrt{2}$ for each walk prepares all quantum walks in the network in the state
\begin{align}
    & C^{k}(0)\ket*{\Psi(0)} = \bigotimes_{u \in V} \bigotimes_{\substack{v \in \delta(u),\\ v > u}} \frac{\ket*{u, c_u} + \ket*{u, c_{uv}}}{\sqrt{2}}.\label{eq:linklevel_start},
\end{align}
where $C^{k}(0) = \bigotimes_{u \in V} C_u$ denotes the coin operator for all $k$ quantum walks.
Applying a flip-flop shift operator following~\eqref{eq:sep_shift} evolves the state in~\eqref{eq:linklevel_start} to the form 
\begin{align}
    & SC^{k}(0)\ket*{\Psi(0)} = \bigotimes_{u \in V} \bigotimes_{\substack{v \in \delta(u),\\ v > u}} \frac{\ket*{u, c_u} + \ket*{v, c_{vu}}}{\sqrt{2}}.\label{eq:linklevel}
\end{align}
Each quantum walk can generate entanglement across data qubits at neighboring nodes with the coin operator in~\eqref{eq:coin_op_assited}.
Note that all quantum walks are in a separable state.

\subsection{Expressing entanglement distribution protocols with QWCP}

Entanglement distribution approaches that rely on path selection for the distribution of Bell states, e.g.,~\cite{chakraborty2020entanglement,li2021effective}, can be expressed using multi-path propagation.
Multipartite entanglement distribution protocols, e.g.,~\cite{meignant2019distributing, bugalho2023distributing}, that use network trees to prepare distributed states can be expressed using tree propagation.
Entanglement distribution protocols that rely on local state knowledge can also be expressed in terms of QWCP~\cite{pant2019routing, patil2022entanglement}.
In this case, the quantum walk operations must be dynamically defined based on local state of the network.
For instance, the coin permutation operation required at an arbitrary node $v$ at instant $t$ may depend on the result of operations performed in the neighbors of $v$ at time $t - 1$.
This dynamic behavior can be obtained based on the exchange of classical information among neighboring nodes in the network, although addressing this in detail is out the scope of this work.

\section{Conclusion}\label{sec:conclusion}

The quantum walk control protocol proposed in this article provides a logical description for a network control plane capable of performing universal distributed quantum computing. The description abstracts the implementation of the quantum walker system and of quantum operations in the network nodes. It considers that quantum error correction yields the application of fault tolerant operators. The key idea that the protocol builds upon is the use of quantum walker systems as quantum control signals that propagate through the network one hop a time. In spite of abstracting physical implementations, the propagation of the walker stipulates latency constraints for the protocol. A generic controlled operation between a qubit in node $A$ with a qubit in node $B$ demands $\mathcal{O}(\Delta(A, B))$ steps of walker evolution. In the context of a possible physical realization of such control system, this latency constraint translates directly to the physical distance between nodes $A$ and $B$. The description of the protocol in the logical setting also masks the effects of walker propagation in the fidelity of distributed operations. When considering imperfect operators, the fidelity of the final outcome is bounded by the fidelity of the coin and shift operators in the quantum walk system.
The protocol described was applied to the task of entanglement distribution and the results highlight connections between the proposed protocol and entanglement distribution protocols.


There are three clear directions for future work considering our results. The first relates to the investigation of physical distributed implementations for quantum walk systems in a network. Implementations of this sort would allow for a realistic characterization of quantities such as fidelity and latency. The second point is the description of control exclusively with quantum information. In this setting, the quantum walk control plane would implement search algorithms in the network graph~\cite{shenvi2003quantum}, substituting the path propagation defined in this work.
Finally, the multiple ways that the quantum walk protocol can be used to implement the same quantum gate in the network allows for the investigation of algorithms for optimizing circuit distribution in quantum networks.



{\em Acknowledgments}---This research was supported in part by the NSF grant CNS-1955744, NSF- ERC Center for Quantum Networks grant EEC-1941583, and MURI ARO Grant W911NF2110325.

\bibliographystyle{IEEEtran}
\bibliography{references}

\end{document}